\documentclass[12pt]{amsart}
\usepackage{amsbsy,amssymb,amscd,amsfonts,latexsym,amstext,delarray,
amsmath,graphicx} 
\input xypic

\usepackage{color}

\newtheorem{thm}{Theorem}[section]
\newtheorem{prop}[thm]{Proposition}

\newtheorem{lem}[thm]{Lemma}

\newtheorem{defn}[thm]{Definition}

\newtheorem{ex}[thm]{Example}

\numberwithin{equation}{section}

\def\bL{{\mathbb L}}

\def\A{{\mathbb A}}
\def\C{{\mathbb C}}
\def\F{{\mathbb F}}

\renewcommand{\P}{{\mathbb P}}
\def\Q{{\mathbb Q}}
\def\R{{\mathbb R}}
\def\Z{{\mathbb Z}}
\def\K{{\mathbb K}}

\def\cI{{\mathcal I}}

\def\cL{{\mathcal L}}

\def\cO{{\mathcal O}}
\def\cP{{\mathcal P}}

\def\cV{{\mathcal V}}

\def\cZ{{\mathcal Z}}

\def\GL{{\rm GL}}

\def\Spec{{\rm Spec}}

\def\Tr{{\rm Tr}}

\title{Motivic Information}
\author{Matilde Marcolli}
\date{}
\address{California Institute of Technology \\ USA \newline \indent
Perimeter Institute for Theoretical Physics \\ Canada \newline \indent
University of Toronto \\ Canada}
\email{matilde@caltech.edu}

\begin{document}
\maketitle

\begin{abstract}
We introduce notions of information/entropy and information loss
associated to exponentiable motivic measures. We show that
they satisfy appropriate analogs to the Khinchin-type properties
that characterize information loss in the context of measures on finite sets. 
\end{abstract}

\begin{center}
{\em In memory of Paolo de Bartolomeis}
\end{center}

\section{Introduction}

I was invited to contribute a paper to a volume of the Bulletin of the
Italian Mathematical Society dedicated to the memory of Paolo de
Bartolomeis.  I met Paolo during my postdoc years at MIT, while 
he was visiting Gang Tian. Since that time, he has always been a nice 
and generous friend, and I regret the fact that we no longer had occasions to see  
each other in recent years: after the main focus of my own research  
shifted away from the area of differential geometry we no longer
frequented the same conferences and the occasions to meet professionally
became much more sporadic. I was deeply saddened by the
news of his untimely death this year. In thinking about a possible contribution
to this volume, I decided to avoid the typically more formal style 
of mathematical papers, which seemed to me a bit too dry for the occasion, 
and I settled instead for a more freely flowing collection of thoughts, somewhat
speculative in nature, revolving around the ideas 
of entropy and information loss, revisited in 
the context of motivic measures.

\smallskip
\subsection{Entropy and information}

The relation between Entropy and Information is one of the fundamental
ideas of contemporary science, introduced by Shannon in the first extensive
mathematical account of the theory of information and communication, \cite{Shannon}.
The Shannon entropy detects the information content of a probability measure
and constrains the amount of information that can be transmitted on a channel,
in terms of a bound on data compression. 
In the simplest case of a probability measure $P=(P_i)$ on a finite set of cardinality $n$,
the Shannon entropy is given by 
\begin{equation}\label{ShannonS}
S(P)=- \sum_{i=1}^n P_i \log P_i .
\end{equation}
There is an axiomatic characterization of the Shannon
entropy given by the Khinchin axioms \cite{Khinchin},
reformulated in a more coincise way by Faddeev \cite{Faddeev}: 
continuity with a maximum at
equidistribution, additivity over subsystems $S(A\cup B)=S(A) + S(B|A)$, 
and expansibility (a compatibility for changing $n$ by 
restriction to the faces of the simplex of probability measures)
suffice to characterize $S(P)$ completely up to a multiplicative
constant $C>0$. 

\smallskip

Recently, the axiomatic characterization of the Shannon entropy was
reinterpreted in modern categorical terms in  \cite{BFL}, \cite{Leinster}, \cite{Mar}, \cite{MarThor}. 
In particular, we are interested here in the notion of information loss for morphisms of
finite sets with probability measures and its axiomatic characterization discussed in \cite{BFL},
which we will review briefly in \S \ref{InfoLossFinSetsSec}.

\smallskip
\subsection{Information loss in the Grothendieck ring of varieties}

Our goal in this paper is to propose an information theoretic point of view in the
context of motivic measures, where we are interested in quantifying phenomena
of ``information loss", associated to morphisms of algebraic varieties. Motivic measures
are meant here as ring homomorphism from the Grothendieck
ring of varieties to various other rings (the integers in the case of the Euler
characteristic, or a polynomial ring in the case of the Poincar\'e polynomial, etc.).
In particular, the motivic Euler characteristic is the ring homomorphism of Gillet--Soul\'e \cite{GilSoul} 
mapping the Grothendieck ring of varieties to the Grothendieck ring of Chow motives.

\smallskip

The structure of the Grothendieck ring of varieties is very subtle, with phenomena such as
the existence of zero-divisors, including the Lefschetz motive, 
\cite{Borisov}, \cite{Martin}, \cite{Poonen} only recently uncovered. Motivic measures can be
seen as ways to probe the structure of the Grothendieck ring, by mapping it to various kinds
of ``Euler characteristic type" invariants. 

\smallskip

Within this general framework we think it is interesting to consider possible notions of
information associated to the evaluation of a motivic measure on a given variety or motive
and information loss associated to morphisms.

\section{Motivic Measures, Integration, and Motivic Information}

\subsection{Hasse-Weil information function}

For a variety $X$ over a finite field $\F_q$, the Hasse--Weil zeta function is given by
the (exponential) generating function for the number of points of $X$ over the
field extensions $\F_{q^m}$, 
$$ Z(X,t)=\exp\left( \sum_{m\geq 1} \frac{\# X(\F_{q^m})}{m} t^m \right). $$
For a variety $X$ defined over $\Z$ with reductions $X_p$ at the primes $p$,
the associated $L$-function is defined as
$$ L(X,s) = \prod_p Z(X_p, p^{-s}). $$

It is convenient to write the Hasse--Weil zeta function in the equivalent form
$$ Z(X,t) = \prod_x (1- t^{\deg(x)})^{-1}, $$
where the product is over the set of closed points of $X$ and $\deg(x)=[k(x):\F_q]$
with $k(x)$ the residue field of the local ring $\cO_{X,x}$ at $x$.
Indeed, by writing $\# X(\F_{q^m}) = \sum_{r |m} r \, a_r$ with $a_r=\# \{ x \,:\, [k(x):\F_q]=r \}$,
one obtains 
$$ Z(X,t)=\prod_{r\geq 1} (1-t^r)^{-a_r}. $$
Equivalently, for $\alpha =\sum_i n_i x_i$ effective zero-cycles 
with $n_i\in \Z_{\geq 0}$ and $x_i$ closed points of $X$, one can write
$$ Z(X,t)=\sum_\alpha t^{\deg(\alpha)}, $$
where $\deg(\alpha)=\sum_i n_i \deg(x_i)$.

It is natural, if one regards the Hasse-Weil zeta function as a motivic measure,
as in \cite{Ram}, \cite{RamTab}, to associate to it an information function of the
form 
\begin{equation}\label{HXt}
 H(X,t):= -\sum_\alpha t^{\deg(\alpha)} \log(t^{\deg(\alpha)}) . 
\end{equation} 
This expression occurs naturally if we write the Shannon entropy
for a distribution of the form
\begin{equation}\label{Palpha}
P(\alpha):= \frac{t^{\deg(\alpha)}}{Z(X,t)}
\end{equation}
over the set of degree zero effective cycles $\alpha$ 
in $X$, that is, the quantity $t^{\deg(\alpha)} / Z(X,t)$ is 
the relative weight assigned by the zeta function to a 
degree zero effective cycle $\alpha$ in $X$.

\begin{defn}\label{localHWS}
For a variety $X$ over a finite field $\F_q$, the local Hasse--Weil entropy is
defined as the Shannon entropy of the distribution $P=(P(\alpha))$
of \eqref{Palpha} on degree zero effective cycles,
\begin{equation}\label{SPp}
 S(X):=-\sum_\alpha P(\alpha) \log(P(\alpha))= 
 \log Z(X,t) + Z(X,t)^{-1} H(X,t) .
\end{equation}
\end{defn}

In the classical Shannon entropy case, for a product distribution $PQ$ one has
$$ S(PQ)=- \sum_i \sum_j P_i Q_j \log(P_i Q_j) 
= -\sum_i P_i \log(P_i) - \sum_j Q_j \log(Q_j), $$
that is, the usual additivity property for independent systems.

Thus, in the case of a variety $X$ over $\Z$ one can consider 
the reductions $X_p$ at the various primes, with the corresponding
Hasse--Weil zeta functions, as independent systems and assign to $X$
an information function of the form 
\begin{equation}\label{HZXs}
 H_\Z(X,s) := \sum_p Z(X_p,p^{-s})^{-1} H(X_p, p^{-s}) . 
\end{equation} 

This corresponds to a distribution $P(\alpha) =\prod_p P(\alpha_p)$ with
\begin{equation}\label{PalphaZ}
 P(\alpha_p)=\frac{p^{-s \deg(\alpha_p)}}{Z(X_p,p^{-s})}. 
\end{equation}

\begin{defn}\label{globalHWS}
For a variety $X$ over $\Z$, the global Hasse--Weil entropy is the
Shannon entropy of the distribution \eqref{PalphaZ},
\begin{equation}\label{SPZ}
 S(X):= \sum_p Z(X_p,p^{-s})^{-1} H(X_p,p^{-s})
+ \sum_p \log Z(X_p,p^{-s}) = H_\Z(X,s) + \log L(x,s).
\end{equation}
\end{defn}

In both \eqref{SPp} and \eqref{SPZ} we see that the Shannon entropy
consists of a term of the form $\log Z(X,t)$ or $\log L(X,s)$ and a term
of the form $H(X,t)$ normalized by the zeta function. In fact, the
Hasse--Weil entropy can be completely described in a simple form in
terms of the logarithm of the arithmetic $L$-function.

\begin{prop}\label{SPZsimple}
The Hasse--Weil entropy \eqref{SPZ} is given by
$$ S(X)=  \log L(X,s) + s \sum_p \log(p) \sum_{m\geq 1} \# X_p(\F_{p^m}) p^{-s m}. $$
The latter term can be equivalently written as $s \frac{d}{ds} \log L(X,s)$, so that
\begin{equation}\label{SPZlog}
S(X) = (1-s \frac{d}{ds}) \log L(X,s).
\end{equation}
\end{prop}

\proof The term $H(X_p, p^{-s})$ is simply 
$$ H(X_p, p^{-s}) = s \log(p) \, \sum_\alpha p^{- s \deg(\alpha)} \deg(\alpha) = 
s \log(p) (t \frac{d}{dt} Z(X_p,t) )|_{t=p^{-s}} $$
$$ Z(X_p,p^{-s})^{-1} H(X_p,p^{-s}) = s \log(p) (t  Z(X_p,t)^{-1} \frac{d}{dt} Z(X_p,t) )|_{t=p^{-s}}. $$
For a generating function $G(t)=\exp(\sum_r c_r \frac{t^r}{r})$ in exponential form,
one has $t \frac{1}{G} \frac{d G}{dt} = t \frac{d \log G}{dt}=\sum_r c_r t^r$. 
This operation corresponds to passing to
ghost components in the Witt ring, as we discuss below. Thus, we obtain 
$$ Z(X_p,p^{-s})^{-1} H(X_p,p^{-s}) = s \log(p) \sum_{m\geq 1} \# X_p(\F_{p^m}) p^{-s m} . $$
We have
$$ \frac{d}{ds} L(X,s) = \frac{d}{ds} \prod_p Z(X_p, p^{-s}) = 
\sum_p \frac{d}{ds} Z(X_p, p^{-s}) \cdot \prod_{\ell \neq p} Z(X_\ell, \ell^{-s}) $$
$$ = \sum_p Z(X_p, p^{-s})^{-1} \frac{d}{ds} Z(X_p, p^{-s}) \cdot L(X,s) = L(X,s) \cdot \sum_p \frac{d}{ds} \log Z(X_p, p^{-s}). $$
This gives
$$ \frac{d}{ds} \log L(X,s) = \sum_p \frac{d}{ds} \log Z(X_p, p^{-s}) =- \sum_p \log(p) t\frac{d}{dt} \log Z(X_p,t) |_{t=p^{-s}}. $$
Thus, we obtain
$$ Z(X_p,p^{-s})^{-1} H(X_p,p^{-s}) =  - s \frac{d}{ds} \log L(X,s)  $$
Thus, we obtain the simpler expression for the Hasse--Weil entropy  of the form \eqref{SPZlog}.
\endproof

The explicit $\log(p)$ factors can be absorbed into a change of basis, using base $p$ logarithm
in the expression for the entropy local factor $H(X_p,p^{-s})$. 

\smallskip
\subsubsection{Hasse-Weil entropy of a point}

\begin{ex}\label{point}
For $X_p={\rm Spec}(\F_p)$ the Hasse-Weil entropy \eqref{SPZ} is given by
\begin{equation}\label{HWSpoint}
S({\rm Spec}(\F_p)) = (1-s\frac{d}{ds}) \log \zeta(s),
\end{equation}
where $\zeta(s)$ is the Riemann zeta function.
\end{ex}

\proof This is immediate from Proposition~\ref{SPZsimple}. It can also be seen 
by direct computation as follows.
For $X_p={\rm Spec}(\F_p)$  we have $Z({\rm Spec}(\F_p),p^{-s})=(1-p^{-s})^{-1}$ 
and $L(X,s)=\zeta(s)=\prod_p (1-p^{-s})^{-1}$. Thus we have
$$ Z({\rm Spec}(\F_p),p^{-s})^{-1} H({\rm Spec}(\F_p),p^{-s}) = \frac{ s \log(p) p^{-s} }{ (1-p^{-s}) } . $$
Thus, in this case the first term in the Shannon entropy \eqref{SPZ} is given by
$$ \sum_p Z(X_p,p^{-s})^{-1} H(X_p,p^{-s}) = s \sum_p \frac{ \log(p) p^{-s} }{ (1-p^{-s}) } 
= s \sum_p \log(p)\sum_{k\geq 1} p^{-ks} = s \sum_n \Lambda(n) n^{-s} $$
where $\Lambda(n)$ is the von Mangoldt function
$$ \Lambda(n) =\left\{ \begin{array}{ll} \log(p) & n =p^k, \,\, k>0 \\ 0 & \text{otherwise.} \end{array}\right. $$
Thus, we have
$$ \sum_p Z(X_p,p^{-s})^{-1} H(X_p,p^{-s}) = - s \, \frac{\zeta^\prime (s)}{\zeta(s)}, $$
where $\zeta(s)$ is the Riemann zeta function. 
The second term in \eqref{SPZ} is simply given by $\log L(X,s) =\log \zeta(s)$.
Thus, the Hasse--Weil entropy in this case is given by
$\log \zeta(s) - s (\log\zeta(s))^\prime$.
\endproof

In Quantum Statistical Mechanics, given a system with partition function
$Z(\beta) =\Tr(e^{-\beta H})$, the entropy can be computed as the
function $S=\frac{\partial}{\partial T} ( T \log Z )$, where $T =1/\beta$ is the
temperature parameter. This is the same as
$$ S = ( 1- \beta \frac{\partial}{\partial \beta})  \log Z(\beta), $$
expressed in terms of the inverse temperature $\beta$. Thus, we
see that the computation of the Hasse--Weil entropy of a point 
given in Lemma~\ref{point} is exactly the thermodynamical entropy of
a quantum statistical mechanical system that has the Riemann
zeta function as partition function. It is well known that the Riemann
zeta function admits an interpretation as partition function in
Quantum Statistical Mechanics, either in terms of the simpler
``Riemann gas" system of \cite{Julia}, \cite{Spector}, 
or in terms of the more refined Bost--Connes system \cite{BC}
(see also \cite{CoMa}). 

\smallskip
\subsubsection{Hasse--Weil entropy of affine spaces} 

\begin{ex}\label{affine}
For $X=\A^n$ the Hasse--Weil entropy is given by
\begin{equation}\label{HWSaffine}
S(\A^n)= \log \zeta(s-n) + s \sum_p \log(p) \frac{p^{-(s-n)}}{1-p^{-(s-n)}} 
=(1-s \frac{d}{ds}) \log \zeta(s-n) . 
\end{equation}
\end{ex}

\proof
For $X=\A^n$ we have $Z(X_{\F_q},t)=(1-q^n t)^{-1}$ and
$L(\A^n,s)=\prod_p (1-p^{-s+n})^{-1}=\zeta(s-n)$. Thus,
the Hasse--Weil entropy is given by \eqref{HWSaffine}.
\endproof

Thus, the effect of passing from a point to an affine space $\A^n$
is simply a shift in the inverse temperature variable $\beta \mapsto \beta -n$
of the quantum statistical mechanical system, namely one obtains
the entropy of a system with partition function $Z_n(\beta)=Z(\beta -n)$.
As $n$ grows large, this system captures the thermodynamical properties 
of the original systems at increasingly low temperatures, that is, for
inverse temperatures $\beta > n$. 

\smallskip
\subsubsection{Hasse--Weil entropy of projective spaces}

\begin{ex}\label{projective}
For $X=\P^n$ the Hasse--Weil entropy is given by
\begin{equation}\label{HWSPn}
S(\P^n)=(1-s\frac{d}{ds}) \prod_{m=0}^n \zeta(s-m).
\end{equation}
\end{ex}

\proof For $X=\P^n$ we have 
$$ Z(\P^n_{\F_q},t)=\frac{1}{(1-t) (1-qt) \cdots (1-q^n t)} $$
hence the $L$-function is given by 
$$ L(\P^n,s) =\prod_{m=0}^n \zeta(s-m). $$
The expression \eqref{HWSPn} is then immediate from Proposition~\ref{SPZsimple}.
\endproof

The expression \eqref{HWSPn} also agrees with the thermodynamical entropy
of a known quantum statistical mechanical system. Indeed, the $\GL_n$ generalizations
of the Bost--Connes system considered in \cite{Shen} (see also the ``determinant part"
considered in \cite{CoMa2}) have partition function $Z(\beta)=\prod_{m=0}^n \zeta(\beta-m)$
and entropy \eqref{HWSPn}.

\subsection{Exponentiable motivic measures and zeta functions}\label{expmeasSec}

The Grothendieck ring $K_0(\cV_\K)$ of varieties over a field $\K$ is generated by isomorphism classes
$[X]$ of varieties with the inclusion-exclusion relation $[X]=[Y]+[X\smallsetminus Y]$ for $Y\subset X$ a
closed subvariety and with the product given by $[X]\cdot [Y]=[X \times Y]$, the class of the product over
$\Spec(\K)$. The Lefschetz motive $\bL=[\A^1]$ is the class of the affine line. 

\smallskip

We follow the terminology used for instance in \cite{RamTab} and we call {\em motivic measure}
any ring homomorphisms $\mu: K_0(\cV_\K) \to R$, where $R$ is a commutative ring. 

\smallskip

When one interprets the classes $[X]$ in the Grothendieck ring as a universal Euler characteristic (see \cite{Bitt})
a motivic measure in the sense specified above is determined by (and in turn determines) an invariant of 
algebraic varieties that satisfies the two main properties of an Euler characteristic, namely inclusion-exclusion
$\mu(X)=\mu(Y)+\mu(X\smallsetminus Y)$ and mulitiplicativity under products $\mu(X\times Y)=\mu(X) \mu(Y)$. 

\smallskip

As shown in \cite{Kap}, \cite{Ram}, \cite{RamTab}, 
to any motivic measure $\mu: K_0(\cV_\K) \to R$ one can associate the Kapranov zeta function, which can be 
seen as a map $\zeta_\mu(\cdot, t): K_0(\cV_\K) \to W(R)$ with values in the big Witt ring $W(R)$ of $R$, and
is defined as
\begin{equation}\label{Kapzeta}
\zeta_\mu(X,t) := \sum_{n=0}^\infty \mu( [S^n(X)] ) \, t^n ,
\end{equation}
where $S^n(X)$ is the $n$-fold symmetric product of $X$, given by the quotient $S^n(X)=X^n/S_n$ of
the $n$-fold product by the action of the symmetric group $S_n$ of permutations.  
This can be regarded as an exponentiated version of the original measure $\mu$, by
interpreting the terms $\mu( [S^n(X)] )$ as analogs of the terms $\mu(X)^n/n!$ in an 
exponential series, \cite{Ram}. 

\smallskip

Here we view the left-hand-side of
\eqref{Kapzeta} as an element in $(1+R[[t]])^*$ and we identify the big Witt ring $W(R)$, as an additive group,
with $((1+R[[t]])^*,\times)$ with the usual product of formal series, which is the addition $+_W$ of the Witt ring,
while the product $\star$ of the Witt ring is uniquely determined by setting
\begin{equation}\label{prodWR}
 (1-at)^{-1} \star (1-bt)^{-1} = (1-ab t)^{-1} 
\end{equation} 
for all $a,b \in R$, see  \cite{Alm}, \cite{Bloch}. In general, the zeta function \eqref{Kapzeta} defines
a group homomorphism $\zeta_\mu(\cdot, t): K_0(\cV_\K) \to W(R)$ but not necessarily a ring homomorphism.

\smallskip

A motivic measure $\mu: K_0(\cV_\K) \to R$  is called {\em exponentiable} (see \cite{RamTab}) if the 
associated Kapranov zeta function $\zeta_\mu(\cdot, t): K_0(\cV_\K) \to W(R)$ is a ring homomorphism,
that is, if the zeta function is itself a motivic measure. 

\smallskip

The motivic measure given by the counting of points over finite fields is exponentiable, \cite{Ram},
and the Gillet--Soul\'e motivic measure of \cite{GilSoul} (the motivic Euler characteristic) $\mu_{GS}:
K_0(\cV_\K) \to K_0({\rm Chow}(\K)_\Q)$ is also exponentiable, \cite{RamTab}. Several
motivic measures that factor through  $\mu_{GS}$, like the topological Euler characteristic, 
the Hodge and Poincar\'e polynomials, are also exponentiable (see \cite{RamTab}), while
the Larsen--Lunts motivic measure \cite{LaLu} is not exponentiable, since as shown in Proposition~4.3
of \cite{RamTab} in the exponentiable case if the zeta functions of two varieties are rational then
the zeta function of the product also is, while the Larsen--Lunts motivic measure provides an
example where zeta functions of curves are rational but the zeta function of a product of two 
positive genus curves is not. 

\smallskip

The exponentiable property of motivic measures is related to $\lambda$-ring structures.
A $\lambda$-ring $R$ is a commutative ring endowed with maps 
$\lambda^n: R \to R$ satisfying $\lambda^0(a)=1$, $\lambda^1(a)=a$ and
$\lambda^n(a+b)=\sum_{i+j=n} \lambda^i(a) \lambda^j(b)$, so that
$\lambda_t(a)=\sum_n \lambda^n(a) t^n$ is a group homomorphism
$\lambda_t: R \to W(R)$. 
Assume that $R$ is a $\lambda$-ring such that the group homomorphism
$\sigma_t: R \to W(R)$ given by $\sigma_t(a) =\lambda_{-t}(a)^{-1}$ (the opposite
$\lambda$-structure) is a ring homomorphism.
Then as shown in \cite{Ram}, \cite{RamTab}, the exponentiable condition for a motivic
measure $\mu: K_0(\cV_\K) \to R$ can be phrased as the property that  
\begin{equation}\label{expmu}
 \mu([S^n(X)]) = \sigma^n(\mu([X])), 
\end{equation} 
where $\sigma_t(a)=\sum_n \sigma^n(a) t^n$.

\smallskip

In the following we will restrict our attention to motivic measures that are exponentiable.

\smallskip
\subsection{A motivic entropy function}\label{motentSec}

Given an exponentiable motivic measure $\mu: K_0(\cV_\K) \to R$ and an associated
motivic zeta function $\zeta_\mu(X,t)$, we consider an associated Shannon type
entropy function, which generalizes the Hasse-Weil entropy described in the
previous sections. By analogy to Definition~\ref{localHWS} we expect an
expression of the form
\begin{equation}\label{Smu}
S_\mu(X) := \log \zeta_\mu(X,t) + \zeta_\mu(X,t)^{-1} H_\mu(X,t),
\end{equation}
where we need to specify more precisely what the terms mean in
the context of motivic zeta functions with values in the Witt ring $W(R)$. 
As in the Hasse--Weil case discussed above, we expect the term
$\zeta_\mu(X,t)^{-1} H_\mu(X,t)$ to take the form of a logarithmic
derivative. 
Thus, a candidate definition for
a motivic entropy of an exponentiable motivic measure $\mu:K_0(\cV_\K)\to R$
would be given by
\begin{equation}\label{Smu2}
S_\mu(X) :=  (1-s\frac{d}{ds})  \log \zeta_\mu(X,\lambda^{-s}),
\end{equation}
where $\lambda$ is a parameter in $\R^*_+$ and the change of variables $t=\lambda^{-s}$ is
meant to interpret the $s$ variable as an inverse temperature thermodynamic parameter. 
This means interpreting the motivic zeta function $\zeta_\mu(X,\lambda^{-s})$
as a partition function and \eqref{Smu2} as its thermodynamical entropy. 

\smallskip

In terms of the $t$ variable, this means defining the entropy function as
\begin{equation}\label{Smu2t}
S_\mu(X) =  (1- t \log (t) \frac{d}{dt})  \log \zeta_\mu(X,t).
\end{equation}

\smallskip
\subsubsection{Lambda ring structure and Adams operations}

The term $t \frac{d}{dt} \log \zeta_\mu(X,t)$ in \eqref{Smu2t} has a natural
interpretation in terms of lambda ring structures and the associated Adams
operations. Indeed, one defines the $n$-th Adams operation $\Psi_n(a)$ on the $\lambda$-ring $R$
as the $n$-th ghost component of the opposite $\lambda$-structure $\sigma_t(a)$, that is, 
\begin{equation}\label{adams}
  t \frac{d}{dt} \log \sigma_t(a) = \psi_t(a) =\sum_{n\geq 1} \Psi_n(a) t^n. 
\end{equation}  
(Here we follow the sign convention as in \cite{Hess} for $\Psi_n(a)$ rather than
as in \cite{Ram}.) 
These are ring homomorphisms $\Psi_n: R\to R$, satisfying $\Psi_n \circ \Psi_m =\Psi_{nm}$.

\smallskip

\begin{lem}\label{MotS}
Let $R$ be a commutative ring with no $\Z$-torsion and with opposite $\lambda$-ring structure $\sigma_t$.
The motivic entropy \eqref{Smu2} of an exponentiable motivic measure $\mu:K_0(\cV_\K)\to R$
is given by
\begin{equation}\label{SmuPsi}
S_\mu(X) = (1-t \log(t) \frac{d}{dt}) \log \sigma_t (\mu([X]) ) = \sum_{n\geq 1} \frac{\Psi_n(\mu([X]))}{n}  t^n 
- \sum_{n\geq 1} \Psi_n(\mu([X])) \, t^n \log(t). 
\end{equation}
\end{lem}

\smallskip
\subsubsection{Motivic entropy of the Euler characteristics}

As shown in \cite{Ram}, the Macdonald formula for the Euler characteristics of
symmetric products 
\begin{equation}\label{chisymm}
\sum_{n=0}^\infty \chi(S^n(X)) t^n = (1-t)^{-\chi(X)} =\exp(\sum_{r>0} \chi(X) \frac{t^r}{r})
\end{equation}
implies that the motivic measure on $K_0(\cV_\C)$ given
by the Euler characteristic can be exponentiated. We can also read directly the value
of the associated entropy function from \eqref{chisymm}. We obtain the following.

\begin{ex}\label{chiSex}
The motivic entropy of the motivic measure $\chi: K_0(\cV_\C) \to \Z$ given by
the Euler characteristics is given by
\begin{equation}\label{Schi}
\begin{array}{rl} 
S_\chi(X) = & \displaystyle{ (1-t \log(t) \frac{d}{dt}) \log (1-t)^{-\chi(X)} } \\[3mm]
 = & \displaystyle{ \chi(X) \frac{S(t,1-t)}{(1-t)} } \\[3mm]
 = & \displaystyle{ \chi(X)\, \zeta_\chi({\rm Spec}(\K),t) \,S(t,1-t) },
 \end{array}
\end{equation}
where $S(t,1-t)=-t \log(t) -(1-t) \log (1-t)$ is the binary Shannon entropy function
and $\zeta_\chi({\rm Spec}(\K),t)=(1-t)^{-1}$ is the zeta function of a point.
\end{ex}

\begin{center}
\begin{figure}[h]
\includegraphics[scale=0.5]{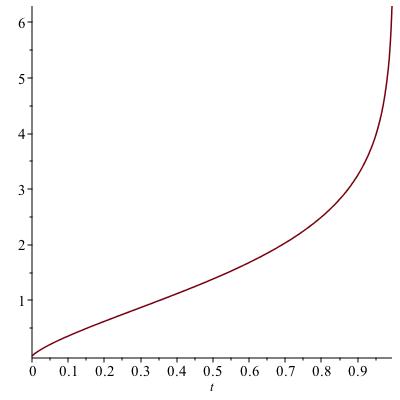} 
\caption{The motivic entropy of the Euler characteristic. \label{chiSt}}
\end{figure}
\end{center}

We should regard the dependence of the entropy on the variable $t$ as a thermodynamic
parameter, namely after a change of variable $t=e^{-\beta}$ we can think of the zeta function
$$ \sum_{n=0}^\infty \chi(S^n(X)) e^{-n \beta} $$
as a partition function, where (at least in the case of non-negative Euler characteristics) 
the coefficient $\chi(S^n(X))$ represents the degeneracy of the $n$-th energy level.
In this view, the behavior of the function \eqref{chisymm} with respect to $t$, shown in
Figure~\ref{chiSt} for a value $\chi(X)=1$, corresponds near $t=0$ (large $\beta\to \infty$)
to the low temperature $T\to 0$ behavior of the system, while the behavior near $t=1$ 
(near $\beta=0$) corresponds to the high temperature $T\to \infty$ limit. 

\smallskip
\subsubsection{Motivic entropy of Poincar\'e polynomials}

Similarly, the Mcdonald formula for the Poincar\'e polynomials, 
\begin{equation}\label{McdonaldP}
\sum_{n=0}^\infty \cP(S^n(X),z) t^n =\prod_{j=0}^{2n} (1-z^j t) ^{(-1)^{j+1} b_j(X)} = \exp(\sum_{r>0} \cP(X,z^r) \frac{t^r}{r} ),
\end{equation}
used in \cite{Ram} to show that the associated motivic measure is exponentiable,
gives the value of the motivic entropy.

\begin{ex}\label{exPoinc}
The motivic entropy of the motivic measure defined by the Poincar\'e polynomial 
is given by
\begin{equation}\label{SPoinc}
S_\cP(X)=  \sum_{j=0}^{2n} (-1)^j b_j(X) \tau(z^j) \, ( S(z^j t, 1-z^j t) + z^j t \log(z^j) ),
\end{equation}
where $\tau: \Z[z]\to W(\Z[z])$ is the Teichm\"uller character to the Witt ring
and $S(u,1-u)=-u\log(u)-(1-u)\log(1-u)$ is the binary Shannon entropy. 
\end{ex}

\proof We have
$$ S_\cP(X)= (1-t \log(t) \frac{d}{dt}) \log \zeta_\cP(X,t) =
(1-t \log(t) \frac{d}{dt})  \sum_{j=0}^{2n} (-1)^{j+1} b_j(X) \log (1-z^j t) $$
$$ = \sum_j (-1)^{j+1} b_j(X) ( \log (1-z^j t) + \frac{z^j t  \log(t)}{ 1-z^j t } ) $$
$$ = \sum_j \frac{(-1)^j b_j(X)}{1-z^j t} (-(1-z^j t)\log (1-z^j t) - z^j t \log(z^j t) + z^j t \log(z^j) ) $$
$$ = \sum_j \frac{(-1)^j b_j(X)}{1-z^j t} (S(z^j t, 1-z^j t)+ z^j t \log(z^j) ), $$
where $(1-z^j t)^{-1} =\tau(z^j)$ is the image in the Witt ring $W(\Z[z])$ of the
element $z^j\in \Z[z]$ under the Teichm\"uller character $\tau: R \to W(R)$ 
mapping $R\ni a \mapsto \tau(a)=(1-at)^{-1} \in W(R)$.
\endproof

Note that the shift in the binary Shannon entropy $S(z^j t, 1-z^j t) + z^j t \log(z^j)$ 
is similar to the shift of the Shannon entropy one usually encounters in coding theory,
where the $q$-ary Shannon entropy is defined as
$$ S_q(\delta,1-\delta) = S(\delta,1-\delta) + \delta \log_q (q-1) =
- \delta \log_q \delta -(1-\delta) \log_q (1-\delta) + \delta \log_q (q-1). $$
This is the form of the Shannon entropy that describes the asymptotic 
behavior of the volume of the Hamming balls (see for instance \cite{TsfaVla}).

\section{Khinchin Properties of Motivic Entropy}

The classical Shannon entropy is characterized in terms
of the Khinchin axioms, \cite{Khinchin}. It is natural to
consider the question of what formal properties, analogous
in some sense to the Khinchin characterization of entropy,
are satisfied by the motivic version described above. 

\smallskip
\subsection{Extensivity of motivic entropy}

The main property of the Shannon entropy is
the extensivity property, namely its additive
behavior on subsystems. The extensivity
property is usually expressed as the relation
$$ S(A\cup B) = S(A) + S(B|A) = S(B) + S(A|B). $$
We show here that
the analogous property satisfied by the
motivic entropy is the inclusion--exclusion
property, where we think of subvarieties of
a given ambient variety as subsystems and
we identify the conditional entropy with the difference
$$ S_\mu(B|A) = S_\mu(B) - S_\mu(A\cap B) . $$
The case of additivity over independent subsystems
then becomes just the scissor-congruence relation
$[X]=[Y]+[X\smallsetminus Y]$ in the Grothendieck
ring inherited by the entropy function $S_\mu$.

\smallskip

\begin{prop}
The motivic entropy $S_\mu(X)$ of an exponentiable
motivic measure $\mu:K_0(\cV_\K) \to R$ satisfies 
\begin{itemize}
\item Additivity over independent subsystems: for closed 
embeddings $Y\hookrightarrow X$
\begin{equation}\label{SmuIE1}
S_\mu(X) = S_\mu(Y) + S_\mu(X\smallsetminus Y).
\end{equation}
\item Extensivity over subsystems: inclusion--exclusion 
\begin{equation}\label{SmuIE2}
S_\mu(X_1 \cup X_2) = S_\mu(X_1) + S_\mu(X_2) - S_\mu(X_1 \cap X_2).
\end{equation}
\end{itemize}
\end{prop}

\proof
A motivic measure $\mu: K_0(\cV_\K) \to R$ is a ring
homomorphism. In particular, the Grothendieck group
relations $[X]=[Y]+[X\smallsetminus Y]$ for closed 
embeddings $Y\hookrightarrow X$ imply that
$\mu(X)=\mu(Y)+\mu(X\smallsetminus Y)$, which in
turn implies the more general inclusion--exclusion
property $\mu(X_1 \cup X_2)=\mu(X_1)+\mu(X_2)-\mu(X_1\cap X_2)$.

\smallskip

The motivic zeta function $\zeta_\mu(X,t)$ in turn satisfies the relation
\begin{equation}\label{zetainclexcl1}
\zeta_\mu(X,t)=\zeta_\mu(Y,t) \zeta_\mu(X\smallsetminus Y,t) =
\zeta_\mu(Y,t) +_W \zeta_\mu(X\smallsetminus Y,t),
\end{equation}
where the addition $+_W$ in the Witt ring is the multiplication
of power series.  More generally, for $X=X_1 \cup X_2$, one has
\begin{equation}\label{zetainclexcl2}
\zeta_\mu(X,t)=\frac{\zeta_\mu(X_1,t) \zeta_\mu(X_2,t)}{\zeta_\mu(X_1\cap X_2,t)} =
\zeta_\mu(X_1,t) +_W \zeta_\mu(X_2,t) -_W \zeta_\mu(X_1\cap X_2,t).
\end{equation}
Thus, the motivic entropy satisfies \eqref{SmuIE2}.
\endproof

\smallskip
\subsection{Mutual motivic information} 

In information theory the mutual information of two
systems is defined as
$$ \cI(X,Y)=S(X) + S(Y) - S(X \cap Y), $$
or equivalently
$$ \cI(X,Y)=\sum_{x,y} P(x,y) \log \frac{P(x,y)}{P(x)P(y)} $$ $$ = - \sum_x P(x)\log P(x) -
\sum_y P(y) \log P(y) + \sum_{x,y} P(x,y) \log P(x,y), $$
which is the expression above. Thus, the mutual information is directly
defined in terms of an inclusion-exclusion form, where one interprets
$\cI(X,Y)$ as the information of $X\cup Y$. 

\smallskip

Thus, in our interpretation of the extensivity of the motivic entropy, 
given two subvarieties $X,Y$ of some ambient variety, we 
can interpret as mutual information the quantity
$$ \cI_\mu(X,Y) =S_\mu (X\cup Y) = S_\mu(X) + S_\mu(Y) - S_\mu(X\cap Y). $$

\smallskip
\subsection{Zeros}

Another of the formal Khinchin properties of the Shannon entropy is the
fact that it is stationary (and in fact maximal) at the uniform distribution
and it is zero at the most non-uniform distributions $P=(P_i)$ where one
of the $P_i=1$ and all others are zero. We discuss here the meaning of
the vanishing of the motivic entropy. 

\smallskip

So far we have treated the motivic entropy function purely formally, without
defining precisely in what ring of functions it is taking values. Because
of the presence of the $\log(t)$ term, we cannot just view this function as
an element of a power series ring $(1+t R[[t]])^*$ or a Witt ring $W(R)$.
It is better to think of $S_\mu(X)$ as an element of a ring $\cL(R,t)$ of 
formal power series of logarithmic type, in the sense of \cite{LoRo}. 

\medskip

We can describe the motivic entropy as follows.

\begin{lem}\label{lemLop}
The motivic entropy $S_\mu$ is the group homomorphism that
fits in the commutative diagram
$$ \xymatrix{ 
R \ar[r]^{\sigma_t \qquad \qquad} & W(R)=(1+tR[[t]] )^*\ar[d]^{\cL} \\
K_0(\cV_\K) \ar[r]^{S_\mu} \ar[u]^{\mu} \ar[ur]^{\zeta_\mu} & \cL(R,t) 
} $$
where $\mu$ is an exponentiable motivic measure, $\sigma_t$ is the
opposite $\lambda$-ring structure, $\cL(R,t)$ is the ring of
formal power series of logarithmic type, and $\cL: W(R) \to \cL(R,t)$ 
$\cL(f)=(1-t\log(t) \frac{d}{dt}) \log(f)$ is a group homomorphism.
\end{lem}

\proof The fact that the composition $\sigma_t \circ \mu=\zeta_\mu$ is
the motivic zeta function is the condition of exponentiability of the
motivic measure $\mu$, see \cite{Ram}, \cite{RamTab}. The map 
homomorphism $\cL(f) =(1-t \log(t) \frac{d}{dt}) \log(f)$ satisfies
the logarithmic functional equation $\cL(f+_W g)=\cL (f\cdot g)=\cL(f)+\cL(g)$,
hence it defines a group homomorphism $\cL: W(R) \to \cL(R,t)$.
\endproof

\smallskip

\begin{lem}\label{kercritL}
The kernel of the motivic entropy $S_\mu$ is the same as the kernel of
the motivic measure $\zeta_\mu$. 
\end{lem}

\proof It suffices to show that the kernel of $\cL$ is trivial.
An element $f\in W(R)=(1+tR[[t]] )^*$ of the form 
$f(t)=\exp(\sum_{n\geq 1} \frac{a_n}{n} t^n )$ 
is in the Kernel of $\cL$ if
$\log(f) =t \log(t) \frac{d}{dt} \log(f)$, which is verified 
as an identity in $\cL(R,t)$ only if $\log(f)=0$,
that is, if $f=1$ is the additive unit of $W(R)$.
Thus, a class $A=\sum_i n_i [X_i]\in K_0(\cV_\K)$ is
in the kernel of $S_\mu$ iff it is in the kernel of the
exponentiated motivic measure, $\zeta_\mu(A)=1$.
\endproof

Thus, we can see the elements $X$ in the kernel of the
motivic measure as corresponding to the distributions with
least information, or in other words they are the source of
information loss in the motivic measure. 

\smallskip
\subsection{Functoriality}

The remaining Khinchin axioms for the Shannon entropy are continuity
over the simplex of measures $P=(P_i)$ and a consistence condition
when viewing an $n$-dimensional simplex as a face of an $(n+1)$-dimensional
simplex, $$S_{n+1}(P_1,\ldots,P_n,0)=S_n(P_1,\ldots,P_n), $$
together with the symmetry of $S$ under permutations of its arguments. 
We can view this requirement as a kind of functoriality requirement, when we consider the
inclusion of faces as morphisms. Thus, the analogous property we
require for the entropy function defined in the motivic setting is to
satisfy a functoriality property induced by the funtoriality of Witt rings.

\begin{lem}\label{functS}
The motivic entropy is functorial. Namely, 
if $\mu: K_0(\cV_\K)\to R$ and $\mu': K_0(\cV_\K)\to R'$ are exponentiable
motivic measures related by a (pre)-$\lambda$-ring homomorphism $\phi: R\to R'$,
so that $\mu'=\phi\circ \mu$, then there exists a group homomorphism $S:\cL(R,t) \to
\cL(R',t)$ such that $S_{\mu'}=S(\phi)\circ S_\mu$.
\end{lem}

\proof
The Witt rings are functorial, in the sense that a ring homomorphism $\phi: R \to R'$
induces a ring homomorphism $W(\phi): W(R) \to W(R')$.  A morphism of (pre)-$\lambda$-rings is
a ring homomorphism $\phi : R\to R'$ for which one has a commutative diagram
$$ \xymatrix{
R \ar[d]^\phi \ar[r]^{\sigma_t} & \Lambda(R) \ar[d]^{\Lambda(\phi)} \\
R' \ar[r]^{\sigma'_t} & \Lambda(R'),
} $$
with $\Lambda(R) =(1+tR[[t]] )^*$.
The ghost map $gh: W(R) \to t R[[t]]$ is also functorial, and so is the ring of formal power 
series of logarithmic type. 
Thus, we obtain a diagram
$$ \xymatrix{
 & R \ar[r]^{\sigma_t} \ar[dd]^{\phi} & \Lambda(R) \ar[dd]^{\Lambda(\phi)} \ar[r]^{\cL} & \cL(R,t) \ar[dd]^{S(\phi)} \\
K_0(\cV_\K) \ar[ur]^{\mu} \ar[dr]^{\mu'} & & &   \\
 & R' \ar[r]^{\sigma'_t} & \Lambda(R') \ar[r]^{\cL} & \cL(R',t).
} $$
\endproof

\medskip
\section{Motivic Entropy as Information Loss} 

The proposal discussed above for a notion of Entropy/Information in the setting
of motivic measures is based on our initial observation that we can interpret the
Hasse--Weil zeta function, when written in terms of effective zero-cycles, as a
distribution as in \eqref{Palpha} for which we formally compute the ordinary Shannon
entropy. The resulting expression was then generalized in the form \eqref{Smu2}
for an arbitrary exponentiable motivic measure. 

\smallskip

This proposal, however, has the drawback that it does not lend itself
easily to a relative form, a motivic version of a Kullback--Leibler 
divergence, or better a measure of information loss associated to
morphisms, which would provide a motivic analog of the
characterization of information loss of \cite{BFL}.

\smallskip

We discuss here how one can modify the original proposal so as
to accommodate a notion of information loss. 

\smallskip
\subsection{Information loss on finite sets}\label{InfoLossFinSetsSec}

In the usual setting of probability measures on finite sets and
classical information theory, given a morphism $f:(\Sigma,P) \to (\Sigma',Q)$,
where $\Sigma,\Sigma'$ are finite sets and $P,Q$ are probability measures,
one counts the information loss of $f$ as a Kullback--Leibler
divergence
\begin{equation}\label{infoloss}
 \cI(f)= S(P)-S(Q)= \sum_{s\in \Sigma} P_s \log\frac{Q_{f(s)}}{P_s} ={\rm KL}(P||Q).
\end{equation} 
The second equality follows by a simple calculation, see \cite{BFL}, using the assumption
that morphisms are measure preserving, namely that
\begin{equation}\label{measpres}
 Q_j = \sum_{i\in f^{-1}(j)} P_i. 
\end{equation} 
In our setting we will need to consider more general morphisms, which do not
necessarily satisfy the condition \eqref{measpres}, hence we will consider the Kullback--Leibler
divergence ${\rm KL}(P||Q)=\sum_{s\in \Sigma} P_s \log\frac{Q_{f(s)}}{P_s}$ 
as our model of information loss, even when this does not necessarily agree with
the difference $S(P)-S(Q)$.

\smallskip

The function $\cI(f)$ of \eqref{infoloss} satisfies an axiomatic characterization 
(up to a constant multiplicative factor), 
which follows from the
Khinchin axioms of the Shannon entropy (reformulated as in \cite{Faddeev}):
\begin{itemize}
\item Additivity under composition of morphisms: 
$\cI(f\circ g)=\cI(f)+\cI(g)$;
\item Additivity under direct sums: $\cI(f\oplus g)=\cI(f)+\cI(g)$;
\item Homogeneity under scaling: $\cI(\lambda f)=\lambda \cI(f)$, for $\lambda\in \R^*_+$.
\end{itemize}
The last two properties are replaced by the single additivity over convex combinations
\begin{equation}\label{convcomb}
\cI(\lambda f\oplus (1-\lambda) g)=\lambda \cI(f)+(1-\lambda) \cI(g), 
\end{equation}
for $\lambda\in [0,1]$,
if the normalization of measures is preserved, see \cite{BFL}.
Additivity under composition plays the role of a functoriality property in
the framework of \cite{BFL}.  

\smallskip
\subsection{Sources of Information Loss}

We are interested here in a similar counting of information loss associated to
motivic measures. As we discussed above, the kernel of an exponentiated 
motivic measure can be viewed as the amount of information contained in the
Grothendieck ring of varieties that is lost when seen through the given
motivic measure. It is also the kernel of the motivic entropy reflecting this
interpretation as information loss. 

\smallskip

If we want to make this idea of information loss in the motivic context
more precise, we can identify two different possible sources of information loss:
\begin{itemize}
\item Ring homomorphisms $\phi: R \to R'$
\item Morphisms of varieties $f: X\to Y$ (or correspondences of motives).
\end{itemize}
The first case corresponds to modifying the motivic measure $\mu: K_0(\cV_\K)\to R$
by composition with a ring homomorphism $\phi: R \to R'$, while keeping the
variety it is evaluated on unchanged, while the second case corresponds to maintaining
the motivic measure unchanged while modifying the varieties through morphisms
 $f: X\to Y$ of algebraic varieties, for motivic measures defined on
the Grothendieck ring of varieties $K_0(\cV_\K)$, or correspondences 
$\alpha: h(X) \to h(Y)$ of Chow motives, for motivic measures on $K_0({\rm Chow}(\K))$.

\smallskip
\subsection{Power structures}

In the next subsection we introduce an information loss function associated
to a triple $(\phi,\mu,\mu')$ consisting of motivic measures
$\mu: K_0(\cV_\K)\to R$ and $\mu': K_0(\cV_\K)\to R'$ and a ring
homomorphism $\phi: R \to R'$.

\smallskip

In order to discuss an analog of the convex combination property
\eqref{convcomb} of information loss, we need to first recall the
notion of a power structure, see \cite{GuZaLuHer}.

\smallskip

\begin{defn}\label{powerdef}
A {\em power structure} on a ring $R$ is a map
$(1+R[[t]]) \times R \to 1 + R[[t]]$, 
$(f(t),a) \mapsto f(t)^a$, with the properties that
\begin{itemize}
\item $f(t)^0=1$, for all $f\in 1 + R[[t]]$,
\item $f(t)^1=f(t)$, for all $f\in 1 + R[[t]]$,
\item $(f(t)\cdot g(t))^a =f(t)^a \cdot g(t)^a$, for all $f,g\in 1 + R[[t]]$, $a\in R$,
\item $f(t)^{a+b}=f(t)^a \cdot f(t)^b$, for all $f\in 1 + R[[t]]$, $a,b\in R$,
\item $f(t)^{ab}=(f(t)^a)^b$, for all $f\in 1 + R[[t]]$, $a,b\in R$.
\end{itemize}
\end{defn}

\smallskip

\begin{ex}
As shown in \cite{GuZaLuHer}, there exists a power structure on the
Grothendieck ring of varieties $K_0(\cV_\C)$ such that the universal
motivic zeta function 
$$ \zeta_{\mu_u}( X,t)=\sum_{n=0}^\infty [S^n(X)] \, t^n, $$
which is the exponentiation of $\mu_u={\rm id}: K_0(\cV_\C)\to K_0(\cV_\C)$,
satisfies 
\begin{equation}\label{powerzeta}
 (1-t)^{-[X]} = \zeta_{\mu_u} ( X,t) .
\end{equation} 
It is obtained by setting
$$ f(t)^{[X]}:= 1 + \sum_{k=1}^\infty  \sum_{\sum i k_i =k} \left[ (\prod_i X^{k_i}\smallsetminus \Delta) \times \prod_i X_i^{k_i} / \prod_i S_{k_i} \right]\, t^k, $$
for $f(t)=1+\sum_i [X_i] \, t^i$ with $[X_i]\in K_0(\cV_\C)$, see \cite{GuZaLuHer} for more details.
\end{ex}

\smallskip
\subsection{Information loss from ring homomorphisms}

A measure of information loss associated to a ring homomorphism
$\phi: R \to R'$ and a pair of given exponentiable motivic measures 
$\mu: K_0(\cV_\K)\to R$ and $\mu': K_0(\cV_\K)\to R'$ can be obtained
simply by the difference of the motivic entropies
\begin{equation}\label{Iphimumuprime}
\cI_X(\phi,\mu,\mu')= S_{\phi\circ \mu}(X) - S_{\mu'}(X) = (1-t\log(t) \frac{d}{dt}) \log \frac{\zeta_{\phi\circ \mu}(X,t)}{\zeta_{\mu'}(X,t)},
\end{equation}
where $S_{\phi\circ \mu}(X) =S(\phi)\circ S_{\mu}(X)$ and
$\zeta_{\phi\circ \mu}(X,t) =\Lambda(\phi) \zeta_{\mu}(X,t)$, 
by Lemma~\ref{functS}. 

\smallskip

This measure of information loss satisfies an analog of the properties of information
loss described in \cite{BFL}.

\begin{lem}\label{RInfo3prop}
Let $\phi: R \to R'$ be a morphism of commutative rings and let 
$\mu: K_0(\cV_\K)\to R$ and $\mu': K_0(\cV_\K)\to R'$ be exponentiable motivic measures.
Then the information loss function $\cI_X(\phi,\mu,\mu')$ of \eqref{Iphimumuprime} satisfies
\begin{enumerate}
\item Additivity under composition $R\stackrel{\psi}{\to} R'\stackrel{\phi}{\to} R''$:
\begin{equation}\label{RinfolossPhi12}
 \cI_X(\phi \circ \psi,\mu,\mu'')=\cI_X(\phi ,\mu',\mu'') + S(\phi) \circ \cI_X(\psi,\mu,\mu').
\end{equation} 
\item Additivity under combination: for $\phi_1,\phi_2: R \to R'$ ring homomorphisms, where
the ring $R'$ has a power structure, 
\begin{equation}\label{addconvex}
 \cI_X(\lambda \phi_1 + (1-\lambda)\phi_2,\mu,\mu')  = \lambda \, \cI_X(\phi_1,\mu,\mu')  +
 (1-\lambda) \, \cI_X(\phi_2,\mu,\mu'),
\end{equation}
where
\begin{equation}\label{Iconvex}
\cI_X(\lambda \phi_1 + (1-\lambda)\phi_2,\mu,\mu')  := (1-t \log(t)\frac{d}{dt}) \log \frac{ \zeta_{\phi_1\circ \mu}(X,t)^\lambda \cdot \zeta_{\phi_2\circ \mu}(X,t)^{1-\lambda} }{\zeta_{\mu'}(X,t)}.
\end{equation}
\end{enumerate}
\end{lem}

\proof For the composition $\phi \circ \psi: R \to R''$, by Lemma~\ref{functS}  we have
$$ S_{(\phi \circ \psi)\circ \mu}(X) - S_{\mu''}(X) =S(\phi \circ \psi) \circ S_\mu(X) - 
S_{\mu''}(X) $$
$$ = S(\phi) \circ S_{\psi \circ \mu}(X) - S(\phi) \circ S_{\mu'}(X) + S_{\phi \circ \mu'}(X) - S_{\mu''}(X) $$
$$ = S(\phi) ( S_{\psi \circ \mu}(X) - S_{\mu'}(X) ) + S_{\phi \circ \mu'}(X) - S_{\mu''}(X), $$
hence we obtain \eqref{RinfolossPhi12}. 

For $\lambda\in R'$, consider the element 
\begin{equation}\label{amalgWR}
\zeta_{(\lambda \phi_1 + (1-\lambda)\phi_2 )\circ \mu}(X,t):=\zeta_{\phi_1\circ \mu}(X,t)^\lambda \cdot \zeta_{\phi_2\circ \mu}(X,t)^{1-\lambda}, 
\end{equation} 
where the product as power series is the addition in the Witt ring and the powers, for 
$\lambda$ and $1-\lambda \in R'$, are determined by the power structure of $R'$, so that
\eqref{amalgWR} is clearly the analog of a convex combination in $W(R')$. 
We have
$$ \cI_X(\lambda \phi_1 + (1-\lambda)\phi_2,\mu,\mu') =(1-t \log(t)\frac{d}{dt}) \log \frac{ \zeta_{(\lambda \phi_1 + (1-\lambda)\phi_2 )\circ \mu}(X,t) }{\zeta_{\mu'}(X,t)} $$
$$ = (1-t \log(t)\frac{d}{dt}) \log \frac{ \zeta_{\phi_1\circ \mu}(X,t)^\lambda \cdot \zeta_{\phi_2\circ \mu}(X,t)^{1-\lambda} }{\zeta_{\mu'}(X,t)^\lambda \cdot \zeta_{\mu'}(X,t)^{1-\lambda}} $$
$$ = \lambda ( S_{\phi_1\circ \mu}(X) - S_{\mu'}(X)) + (1-\lambda) (S_{\phi_2\circ \mu}(X) - S_{\mu'}(X)), $$
so that we obtain \eqref{addconvex}.
\endproof

\smallskip
\subsection{Hasse--Weil information loss}

We then consider the question of how to construct an information loss function
associated to morphisms of varieties. To this purpose we analyze again the
case of the Hasse-Wil zeta function and the motivic
measure given by the counting measure for varieties over finite fields. 

\smallskip

As we have seen before, when we describe the Hasse-Weil zeta function
as a generating function for effective $0$-cycles, we can associate to it
the distribution $P(\alpha) = t^{\deg(\alpha)}/ Z(X,t)$, for $\alpha=\sum_i n_i x_i$
a $0$-cycle on $X$, with $\deg(\alpha)=\sum_i n_i \deg(x_i)$. 

\smallskip

Using the Kullback--Leibler divergence point of view on how to measure information loss,
we aim at computing a relative entropy of the distribution $P=(P(\alpha))$ on $0$-cycles
on $X$ and the corresponding distribution for $0$-cycles on $Y$, by comparing them
via the morphism $f: X\to Y$. 

\smallskip

Cycles push forward under proper morphisms and pull back under flat morphisms. Thus, we
can consider two different information loss functions for these two classes of morphisms. 

\smallskip
\subsubsection{Hasse-Weil information loss for proper morphisms}

Given a proper morphism $f: X \to Y$ of algebraic varieties, for a subvariety
$V\subset X$, one defines the pushforward $f_*(V)$ as zero if $\dim f(V)<\dim V$
and as $f_*(V)=\deg(V/f(V)) \, f(V)$ if $\dim f(V)=\dim V$, where
$\deg(V/f(V))$ is the degree $[\K(V): \K(f(V))]$ of the finite field extension
$\K(V)$ of $\K(f(V))$. The definition is then extended by linearity to combinations
$\sum_i n_i V_i$. In particular, for a $0$-cycle $\alpha =\sum_i n_i x_i$ in $X$,
the pushforward under a proper morphism $f: X\to Y$ is given by
\begin{equation}\label{pushalpha}
f_*(\alpha) = \sum_i n_i \, \deg(x_i/f(x_i)) \, \deg(f(x_i)), 
\end{equation}
where $\deg(x/f(x))=[\K(x):\K(f(x))]$.

\smallskip

Over the field of complex numbers the degree $\deg(x/f(x))$ represents geometrically
the number of points of the fiber $\# f^{-1}(y)$ for $y=f(x)$ (counted with the appropriate
multiplicity in the case of ramification). However, this is not
necessarily the case in positive characteristics, where 
for example the map induced by $\K[t^p]\to \K[t]$ has degree 
$p$ but is one-to-one on points. 

\smallskip

\begin{defn}\label{defHWinfoloss}
The Hasse--Weil information loss of a proper morphism $f:X\to Y$ is given by
\begin{equation}\label{HWinfoloss}
\cI_{HW}(f_*) := \sum_{\alpha \in \cZ^0_{{\rm eff}}(X)} P(\alpha) \log \frac{Q(f_*(\alpha))}{P(\alpha)},
\end{equation}
where $P(\alpha)$ is defined as in \eqref{Palpha}, $\cZ^0_{{\rm eff}}(X)$ is the
set of zero-dimensional effective cycles on $X$, and $Q$ is the analogous 
distribution on $Y$,
$$ Q(\gamma)=\frac{t^{\deg(\gamma)}}{Z(Y,t)}, \ \ \ \text{for} \ \ \gamma \in \cZ^0_{{\rm eff}}(Y). $$
\end{defn}

\smallskip
\subsubsection{Hasse-Weil information loss for flat morphisms} 
Let $f: X \to Y$ be a flat morphism of relative dimension $n$. For an irreducible subvariety $V\subset Y$
the pullback $f^*(V)$ is defined as the $f^{-1}(V)$ and extended by linearity.

\begin{defn}\label{defHWinfolossFlat}
The Hasse--Weil information loss of a flat morphism $f:X\to Y$ is given by
\begin{equation}\label{HWinfolossflat}
\cI_{HW}(f^*) := \sum_{\gamma \in \cZ^0_{{\rm eff}}(Y)} Q(\gamma) \log \frac{P(f^*(\gamma))}{Q(\gamma)}.
\end{equation}
\end{defn}

\smallskip
\subsection{Proper morphisms}

The case of proper morphisms, defined in \eqref{HWinfoloss}, is the one that most closely
resembles the definition of information loss for finite sets that we recalled above from \cite{BFL}.
However, because of the behavior of degrees of cycles under pushfoward, it turns out that
the information loss function $\cI_{HW}(f_*)$ of Definition~\ref{defHWinfoloss} is simply 
a logarithmic difference of zeta function. 

\smallskip

\begin{lem}\label{HWzetaH}
The Hasse--Weil information loss \eqref{HWinfoloss} is given by
\begin{equation}\label{HWlogzetas}
\cI_{HW}(f_*) = \log\frac{Z(X,t)}{Z(Y,t)} .
\end{equation}
\end{lem}

\proof By proceeding as in our previous discussion of the Hasse-Weil entropy, we can equivalently 
write the expression \eqref{HWinfoloss} as
\begin{equation}\label{HWlogzetaH}
\cI_{HW}(f_*) = \log\frac{Z(X,t)}{Z(Y,t)} - Z(X,t)^{-1} H(f_*,t),
\end{equation}
where the term $H(f,t)$ is given by
\begin{equation}\label{Hft}
H(f_*,t) =- \sum_{\alpha\in \cZ^0_{{\rm eff}}(X)} t^{\deg(\alpha)} \log(t^{\deg(f_*(\alpha))-\deg(\alpha)}).
\end{equation}
We have $\deg(x)=[\K(x):\K]$ and similarly $\deg(f(x))=[\K(f(x)):\K]$,
hence these degrees are related by 
$$\deg(x)=[\K(x):\K]=[\K(x):\K(f(x))]\cdot [\K(f(x)):\K] =\deg(x/f(x)) \cdot \deg(f(x)), $$
hence $\deg(f_*(\alpha))=\sum_i n_i d_f(x_i) \deg(f(x_i)) = \sum_i n_i \deg(x_i) =\deg(\alpha)$.
Thus, the term $H(f_*,t)$ of \eqref{Hft} vanishes and one is left with \eqref{HWlogzetas}.
\endproof

We check that this notion of information loss satisfies properties of additivity under composition
and combination.
In order to formulate the appropriate condition of additivity under combination, we consider
a decomposition $X = X_1 \cup X_2$ as a disjoint union, and a corresponding
decomposition $Y=Y_1 \cup Y_2$ with the property that $f_i=f|_{X_i}: X_i \to Y_i$.
We write $f=f_1\oplus f_2$ to refer to such data.
We generalize this to weighted combinations $\lambda f_1\oplus (1-\lambda) f_2$, 
by considering the distribution, for $\alpha=(\alpha_1,\alpha_2)$ with 
$\alpha_i\in \cZ^0_{{\rm eff}}(X_i)$, 
\begin{equation}\label{Qlambda}
 Q_\lambda(\alpha) = Q((\lambda f_1\oplus (1-\lambda) f_2)_*(\alpha) := 
 Q_1((f_1)_*(\alpha_1))^\lambda \cdot Q_2((f_2)_*(\alpha_2))^{1-\lambda} ,
 \end{equation}
where for $\gamma_i \in  \cZ^0_{{\rm eff}}(Y_i)$, we have
$Q_i(\gamma_i) := t^{\deg(\gamma_i)}/Z(Y_i,t)$.
Similarly, we also consider the distribution
 $P_i(\alpha_i) := t^{\deg(\alpha_i)}/Z(X_i,t)$ and the distribution
\begin{equation}\label{Plambda}
P_\lambda (\alpha)= P_1(\alpha_1)^\lambda \cdot P_2(\alpha_2)^{1-\lambda}.
\end{equation}

\begin{prop}\label{addproper}
The Hasse--Weil information loss \eqref{HWinfoloss} satisfies additivity under composition
$$ \cI_{HW}((g \circ f)_*)= \cI_{HW}(f_*)  + \cI_{HW}(g_*) $$
and additivity under combination
$$ \cI_{HW}((\lambda f_1 \oplus (1-\lambda) f_2))_*)= \lambda \cI_{HW}((f_1)_*)  + (1-\lambda) 
\cI_{HW}((f_2)_*). $$
\end{prop}

\proof
Clearly the function $\cI_{HW}(f_*)$ of \eqref{HWlogzetaH} satisfies additivity
under composition since
$$ \cI_{HW}((g \circ f)_*)=\log\frac{Z(X,t)}{Z(W,t)} = \log\frac{Z(X,t)}{Z(Y,t)} + \log\frac{Z(Y,t)}{Z(W,t)} =
\cI_{HW}(f_*)  + \cI_{HW}(g_*) $$
for proper morphisms $f: X \to Y$ and $g: Y\to W$. 

For a decomposition $f_i: X_i \to Y_i$ and $f=f_1\oplus f_2$ as above, we have
$$ \cI_{HW}((f_i)_*) = \log \frac{Z(X_i,t)}{Z(Y_i,t)}. $$
Since $Z(X,t)=Z(X_1,t)\cdot Z(X_2,t)$ and $Z(Y,t)=Z(Y_1,t)\cdot Z(Y_2,t)$, we
have additivity
$$ \cI_{HW}(f_*) =\log \frac{Z(X,t)}{Z(Y,t)} = \log \frac{Z(X_1,t)}{Z(Y_1,t)} + \log \frac{Z(X_2,t)}{Z(Y_2,t)}
= \cI_{HW}((f_1)_*) + \cI_{HW}((f_2)_*) . $$

In the case of weighted combinations the information loss is computed by the Kullback-Leibler divergence
\begin{equation}\label{PlambdaKL}
\sum_\alpha P_\lambda (\alpha) \log \frac{Q_\lambda(\alpha)}{P_\lambda(\alpha)},
\end{equation}
where
$$  Q_\lambda((\alpha_1,\alpha_2)) =
\frac{t^{\lambda \deg((f_1)_*(\alpha_1))}}{Z(Y_1,t)^\lambda} \cdot \frac{t^{(1-\lambda) \deg((f_2)_*(\alpha_2))}}{Z(Y_2,t)^{1-\lambda}}  $$
Arguing as in Lemma~\ref{HWzetaH} above, we see that this gives
$$ \cI_{HW}((\lambda f_1 \oplus (1-\lambda) f_2))_*)= \log \frac{Z(X_1,t)^\lambda \cdot 
Z(X_2,t)^{1-\lambda}}{Z(Y_1,t)^\lambda \cdot Z(Y_2,t)^{1-\lambda}}, $$
which gives the additivity property.
\endproof

\smallskip
\subsection{Finite surjective flat morphisms}
We consider then the case of flat morphisms and we focus on the simpler case of 
finite flat surjective morphisms $f: X\to Y$ of smooth quasi-projective varieties,
with constant degree $\delta=\deg(f)$.
In this case the pullback of effective zero-cycles is given by 
$f^*(\gamma) = \sum_i n_i \sum_{x_{i,j} \in f^{-1}(y_i)} x_{i,j}$,
for $\gamma =\sum_i n_i y_i$ an effective zero-cycle in $Y$, 
with $\deg(f^*(\gamma))=\deg(f)\cdot \deg(\gamma)$. 

\smallskip

\begin{lem}\label{finmorlem}
Let $f: X\to Y$ be a finite flat surjective morphism, with constant degree $\delta=\deg(f)$.
Then the information loss function $\cI_{HW}(f^*)$ of \eqref{HWinfolossflat} is
given by
\begin{equation}\label{HWflat}
\cI_{HW}(f^*) = \log \frac{Z(Y,t)}{Z(X,t)} + (\delta-1)\, t \log(t) \frac{d}{dt} \log Z(Y,t).
\end{equation}
\end{lem}

\proof We have
$$ \sum_\gamma t^{\deg(\gamma)}{Z(Y,t)} \log \frac{Z(Y,t)}{t^{\deg(\gamma)}}
\frac{t^{\deg(f^*(\gamma))}}{Z(X,t)} $$
$$ = \log \frac{Z(Y,t)}{Z(X,t)} - \sum_\gamma t^{\deg(\gamma)} \log t^{(\deg(f)-1) \deg(\gamma)}. $$
As in \S \ref{motentSec} we see that this equals \eqref{HWflat}.
\endproof

\smallskip

We can use this description of the information loss function to give a
more general definition for an arbitrary exponentiable motivic measure.

\begin{defn}\label{flatinfolossmudef}
Let $\mu: K_0(\cV_\K) \to R$ be an exponentiable motivic measure
and let $f: X\to Y$ be a finite flat surjective morphism, with constant degree $\delta=\deg(f)$.
The information loss is given by
\begin{equation}\label{infolossflatmu}
\cI_\mu(f^*):= \log \frac{\zeta_\mu(Y,t)}{\zeta_\mu(X,t)} + (\delta-1) t \log(t) \frac{d}{dt} \log \zeta_\mu(Y,t). 
\end{equation}
\end{defn}

\smallskip
\subsection{Information loss of the Euler characteristics}

We consider again the example of the motivic measure given by
the Euler characteristics. 

\begin{prop}\label{Eulcharinfoloss}
For $\K=\C$ and $\chi: K_0(\cV_\C) \to \Z$ the Euler characteristics, the information
loss of a finite flat surjective morphism $f:X\to Y$ of degree $\delta=\deg(f)$ is given by
\begin{equation}\label{IchiFlat}
\cI_\chi(f^*) = S_\chi(Y) - S_\chi(X) + (\chi(f^{-1}(S))-\delta\cdot \chi(S)) \,\zeta_\chi({\rm Spec}(\K),t)\, t\log(t)
\end{equation}
where $S_\chi(X)$ is the motivic information of the Euler characteristics
as in \eqref{Schi} and $S\subset Y$ is the locus such that $f$ is \'etale over $Y\smallsetminus S$. 
If the morphism $f:X\to Y$ is \'etale, then $\cI_\chi(f^*) = S_\chi(Y) - S_\chi(X)$.
\end{prop}

\proof
By the Macdonald formula we have $\zeta_\chi(X,t)=(1-t)^{-\chi(X)}$. Thus,
we obtain
$$ \cI_\chi(f^*) = \log \frac{(1-t)^{-\chi(Y)}}{(1-t)^{-\chi(X)}} + (\delta-1) t \log(t) \frac{d}{dt} \log (1-t)^{-\chi(Y)} $$
$$ =\frac{-1}{1-t} \left( (\chi(Y)-\chi(X)) (1-t) \log(1-t) - (\delta\cdot \chi(Y) -\chi(Y)) t \log(t) \right) . $$
For a finite flat surjective morphism $f:X\to Y$ with degree $\delta=\deg(f)$, the Euler characteristics satisfies the
Riemann--Hurwitz relation
$$ \chi(X) = \delta \cdot \chi(Y) +\chi( f^{-1}(S) ) - \delta\cdot \chi(S), $$
where $f$ is \'etale over $Y\smallsetminus S$. 
Thus, we can write the above as
$$ \cI_\chi(f^*) = \frac{S(t,1-t)}{1-t} (\chi(Y)-\chi(X)) + (\chi(f^{-1}(S))-\delta \cdot \chi(S)) \frac{t \log(t)}{1-t} $$
$$ = \zeta_\chi({\rm Spec}(\K),t) \,\left( (\chi(X)-\chi(Y)) S(t,1-t) + (\chi(f^{-1}(S))-\delta \cdot \chi(S)) \, t\log(t)\right). $$
In the case where the morphism $f: X\to Y$ is \'etale, we have $\chi(X) = \delta \cdot \chi(Y)$ and we obtain
simply the difference of the entropies
$$ \cI_\chi(f^*) = \zeta_\chi({\rm Spec}(\K),t) \, (\chi(Y)-\chi(X))\, S(t,1-t) = S_\chi(Y) - S_\chi(X). $$
\endproof

\smallskip

In the case of the Euler characteristics, the class of \'etale coverings appears to be
the suitable class of morphisms for which the information loss function behaves as
in the case of finite sets and agrees with the difference of entropies. However, this
is not necessarily the case for arbitrary motivic measures. Indeed, unlike the case of 
Zariski locally trivial fibrations, in general if $f: X\to Y$ is an \'etale covering,
the class $[X]$ in the Grothendieck ring does not necessarily factor as a multiple 
of the class $[Y]$. Indeed, by \cite{LaLu} in characteristic zero the quotient of the 
Grothendieck ring by imposing the relation $[X]=\delta \cdot [Y]$ for \'etale coverings 
of degree $\delta$ is isomorphic to $\Z$ via the Euler characteristics. Thus, one does not expect
in general to have $\cI_\mu(f^*)=S_\mu(Y)-S_\mu(X)$ for \'etale coverings for
an arbitrary motivic measure $\mu$. 

\smallskip
\subsection{Additivity properties}

For a decomposition $X=X_1\cup X_2$ and $Y=Y_1\cup Y_2$ with $f_i=f|_{X_i} : X_i \to Y_i$,
and an exponentiable motivic measure $\mu: K_0(\cV_\K) \to R$ where $R$ has a power structure,
we consider the information loss function
\begin{equation}\label{flatmix}
 \begin{array}{rl}
\cI_\mu( (\lambda f_1 \oplus (1-\lambda) f_2)^*) = & \displaystyle{ \log \frac{\zeta_\mu(Y_1,t)^\lambda \cdot \zeta_\mu(Y_2,t)^{1-\lambda}}{\zeta_\mu(X_1,t)^\lambda \cdot \zeta_\mu(X_2,t)^{1-\lambda}} } \\[3mm]
- &\displaystyle{ (\deg(f)-1) t \log(t) \frac{d}{dt} \log( \zeta_\mu(Y_1,t)^\lambda \cdot \zeta_\mu(Y_2,t)^{1-\lambda}) }. \end{array} 
\end{equation}
In the Hasse-Weil case, this corresponds to considering the distributions
$$ P_\lambda(\gamma)=P_1(f_1^*(\gamma_1))^\lambda P_2(f_2^*(\gamma_2))^{1-\lambda} \ \ \ \text{ and } \ \ \
 Q_\lambda(\gamma) = Q_1(\gamma_1)^\lambda Q_2 (\gamma_2)^{1-\lambda}, $$
with $\gamma=(\gamma_1, \gamma_2)$ with $\gamma_i \in \cZ^0_{{\rm eff}}(Y_i)$ and computing the
Kullback--Leibler divergence
$$ \sum_\gamma Q_\lambda(\gamma) \log \frac{P_\lambda(\gamma)}{Q_\lambda(\gamma)}. $$
Since $\deg(f)=\deg(f_i)$ the information loss \eqref{flatmix} satisfies the additivity property
$$ \cI_\mu( (\lambda f_1 \oplus (1-\lambda) f_2)^*) = \lambda \cI_\mu(f_1^*) + (1-\lambda) \cI_\mu(f_2^*). $$

\smallskip

The question of additivity under composition of morphisms is more delicate, because of
the observation mentioned at the end of the previous subsection on the behavior under
\'etale coverings (and more generally under flat surjective morphisms of constant degree).
A simple example where one recovers the behavior of information loss for finite sets is
given by the following class of varieties and morphisms. 

\smallskip

\begin{ex}\label{0diminfoloss}
Given a variety $Y$ over $\K$ consider the set of $X=Y \times S$ where $S$ is a zero-dimensional
variety of the form $S=\Spec (\oplus_{i=1}^N \K)$, for some $N$. Let $\pi_S: X \to Y$ be the
projection map $\pi_S (s,y)=y$. For this set of varieties and maps the information loss satisfies
\begin{equation}\label{infolossdiff0d}
\cI_\mu(\pi_S^*) =S_\mu(Y) - S_\mu(X).
\end{equation}
In particular, $\cI_\mu(\pi_S^*)$ satisfies both additivity under composition
$\cI_\mu((\pi_S \circ \pi_{S'})^*) =\cI_\mu(\pi_S^*)+\cI_\mu(\pi_{S'}^*)$ and
additivity under combination \eqref{flatmix}.
\end{ex}

\proof For an exponentiable measure $\mu: K_0(\cV_\K) \to R$, the zeta function of a product
satisfies $\zeta_\mu(X,t) = \zeta(Y,t) \star_{W(R)} Z(S,t)$, where $\star_{W(R)}$ is the product 
in the Witt ring. Moreover, since $S$ is a union of $N$ copies of ${\rm Spec}(\K)$ we have
$\zeta_\mu(S,t)=(1-t)^{-N}=(1-t)^{-1} +_{W(R)}\cdots +_{W(R)} (1-t)^{-1}$. Thus, since $(1-t)^{-1}$ is the multiplicative unit of $W(R)$, we obtain
$$ \zeta_\mu(X,t) = \zeta(Y,t) \star_{W(R)} ((1-t)^{-1} +_{W(R)}\cdots +_{W(R)} (1-t)^{-1}) $$ 
$$ = \zeta(Y,t) +_{W(R)}\cdots +_{W(R)} \zeta(Y,t) =\zeta_\mu (Y,t)^N. $$
Thus, we have
$$ \cI_\mu(\pi_S^*) = \log \frac{\zeta_\mu(Y,t)}{\zeta_\mu(X,t)} + (N-1) t \log(t) \frac{d}{dt} \log \zeta_\mu(Y,t) $$
$$ = (1-t\log(t) \frac{d}{dt}) \log \zeta_\mu (Y,t) - \log \zeta_\mu(X,t) + N t \log(t) \frac{d}{dt} \log \zeta_\mu(Y,t) $$
$$ = (1-t\log(t) \frac{d}{dt}) \log \zeta_\mu (Y,t) - (1-t\log(t) \frac{d}{dt}) \log \zeta_\mu (X,t). $$
It is then clear that this difference satisfies the required additivity properties. 
\endproof

\end{document}